\documentclass[a4paper,11pt]{article}
\pdfoutput=1

\usepackage{jcappub}

\usepackage[T1]{fontenc}
\title{\boldmath Measurement of the cosmic muon annual and diurnal flux variation with the COSINE-100 detector}

\collaborationImg{\includegraphics[width=22mm,scale=1] {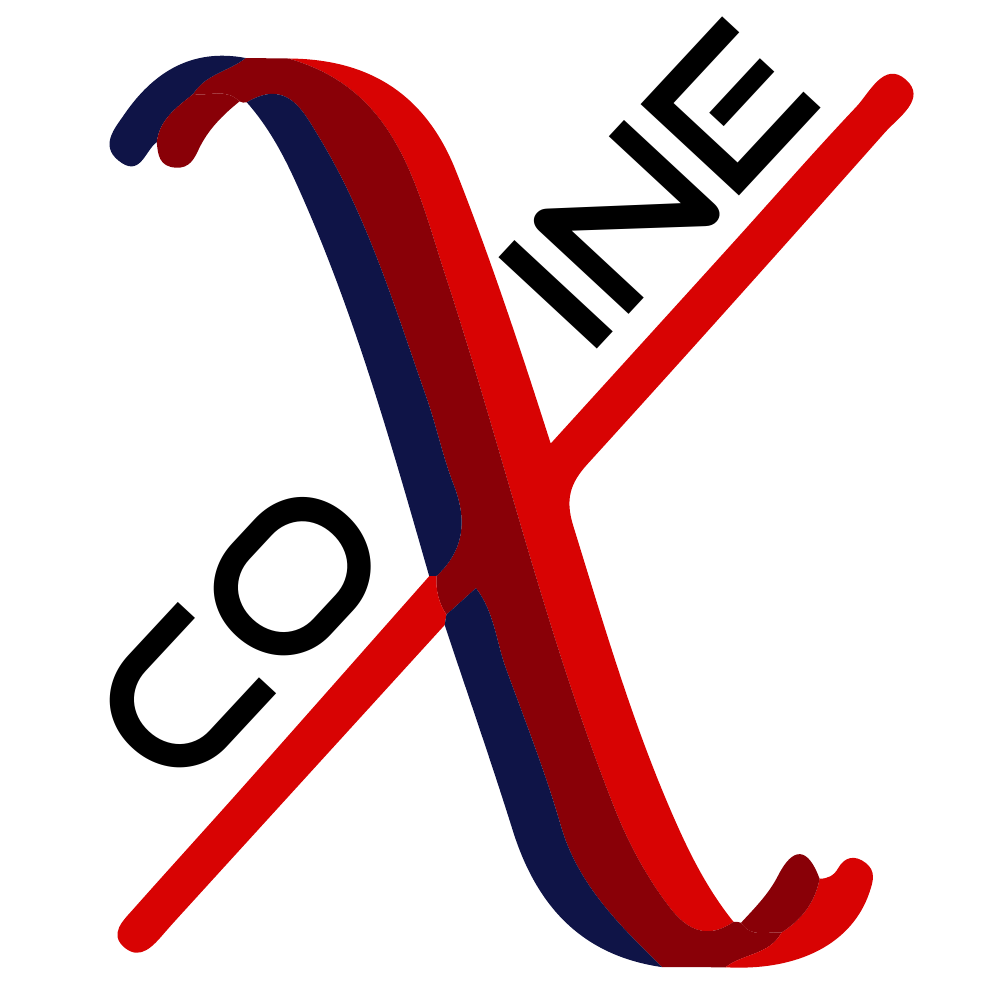}}
\author[a,b]{H.~Prihtiadi,}
\author[c,1]{G.~Adhikari,\note{Present address: Department of Physics, University of California San Diego, La Jolla, CA 92093, USA}}
\author[d]{E.~Barbosa de Souza,}
\author[e]{N.~Carlin,}
\author[f]{J.J.~Choi,}
\author[f]{S.~Choi,}
\author[a]{M.~Djamal,}
\author[g]{A.C.~Ezeribe,}
\author[e]{L.~E.~Fran{\c c}a,}
\author[h]{C.~Ha,}
\author[i]{I.S.~Hahn,}
\author[b]{E.J.~Jeon,}
\author[d]{J.H.~Jo,}
\author[b]{W.G.~Kang,}
\author[j]{M.~Kauer,}
\author[b]{H.~Kim,}
\author[k]{H.J.~Kim,}
\author[b]{K.W.~Kim,}
\author[f]{S.K.~Kim,}
\author[b,c,l]{Y.D.~Kim,}
\author[b,l,m]{Y.H.~Kim,}
\author[b]{Y.J.~Ko,}
\author[b]{E.K.~Lee,}
\author[b,l]{H.S.~Lee,}
\author[b]{J.~Lee,}
\author[k]{J.Y.~Lee,}
\author[b,l]{M.H.~Lee,}
\author[b,l]{S.H.~Lee,}
\author[b]{D.S.~Leonard,}
\author[e]{B.B.~Manzato,}
\author[d]{R.H.~Maruyama,}
\author[g]{R.J.~Neal,}
\author[b]{S.L.~Olsen,}
\author[b,l]{B.J.~Park,}
\author[n]{H.K.~Park,}
\author[m]{H.S.~Park,}
\author[b]{K.S.~Park,}
\author[e]{R.L.C~Pitta,}
\author[b]{S.J.~Ra,}
\author[o]{C.~Rott,}
\author[b]{K.A.~Shin,}
\author[g]{A.~Scraff,}
\author[g]{N.J.C.~Spooner,}
\author[d]{W.G.~Thompson,}
\author[p]{L.~Yang,}
\author[o]{~and G.H.~Yu}

\affiliation[a]{Department of Physics, Bandung Institute of Technology, Bandung 40132, Indonesia}
\affiliation[b]{Center for Underground Physics, Institute for Basic Science (IBS), Daejeon 34126, Republic of Korea}
\affiliation[c]{Department of Physics, Sejong University, Seoul 05006, Republic of Korea}
\affiliation[d]{Department of Physics and Wright Laboratory, Yale University, New Haven, CT 06520, U.S.A.}
\affiliation[e]{Physics Institute, University of S\~{a}o Paulo, S\~{a}o Paulo 05508-090, Brazil}
\affiliation[f]{Department of Physics and Astronomy, Seoul National University, Seoul 08826, Republic of Korea}
\affiliation[g]{Department of Physics and Astronomy, University of Sheffield, Sheffield S3 7RH, United Kingdom}
\affiliation[h]{Department of Physics, Chung-Ang University, Seoul 06973, Republic of Korea}
\affiliation[i]{Department of Science Education, Ewha Womans University, Seoul 03760, Republic of Korea}
\affiliation[j]{Department of Physics and Wisconsin IceCube Particle Astrophysics Center, University of Wisconsin-Madison, Madison, Wisconsin 53706, U.S.A.}
\affiliation[k]{Department of Physics, Kyungpook National University, Daegu 41556, Republic of Korea}
\affiliation[l]{IBS School, University of Science and Technology (UST), Daejeon, 34113, Republic of Korea}
\affiliation[m]{Korea Research Institute of Standards and Science, Daejeon 34113, Republic of Korea}
\affiliation[n]{Department of Accelerator Science, Korea University, Sejong 30019, Republic of Korea}
\affiliation[o]{Department of Physics, Sungkyunkwan University, Suwon 16419, Republic of Korea}
\affiliation[p]{Department of Physics, University of California San Diego, La Jolla, CA 92093, U.S.A.}

\emailAdd{hafizh@ibs.re.kr}
\emailAdd{hyunsulee@ibs.re.kr}

\abstract{We report measurements of annual and diurnal modulations of the cosmic-ray muon rate in the Yangyang underground laboratory (Y2L) using 952~days of COSINE-100 data acquired between September 2016 and July 2019. A correlation of the muon rate with the atmospheric temperature is observed and its amplitude on the muon rate is determined. The effective atmospheric temperature and muon rate variations are positively correlated  with a measured effective temperature coefficient of $\alpha_{T}$~=~0.80 $\pm$ 0.11. This result is consistent with a model of meson production in the atmosphere. We also searched for a diurnal modulation in the underground muon rate by comparing one-hour intervals. No significant diurnal modulation of the muon rate was observed.}


\arxivnumber{----.----}

\begin{document}
\maketitle
\flushbottom

\section{Introduction}
\label{sec:intro}
Although numerous astronomical observations support the conclusion that most of the matter in the universe is invisible dark matter, an understanding of its nature and interactions remains elusive~\cite{Clowe:2006eq,Aghanim:2018eyx}.  The dark matter phenomenon can be attributed to new particles, such as weakly interacting massive particles~(WIMPs)~\cite{PhysRevLett.39.165,Goodman:1984dc} that are well motivated by the theory of supersymmetry~\cite{Jungman:1995df}. Even though tremendous efforts have been pursued to search for WIMP dark matter by directly detecting nuclei recoiling from WIMP-nucleus interactions, no definitive signal has been observed~\cite{Undagoitia:2015gya,Schumann:2019eaa}. One exception is the DAMA/LIBRA experiment that uses an array of NaI(Tl) detectors~\cite{Bernabei:2013xsa,Bernabei:2018yyw} and sees an annual event rate modulation that can be interpreted as being due to WIMP-nuclei interactions~\cite{Savage:2008er}. However, this observation has been the subject of tension since the WIMP-nucleon cross sections inferred from the DAMA/LIBRA modulation are excluded by other experiments~\cite{Tanabashi:2018oca,Aprile:2018dbl}. These observations motivate considerations of environmental effects such as cosmic-ray muons as a possible source of the annual modulation signal~\cite{Nygren:2011,Davis:2014cja}. 

Muons are produced in the decay of mesons created by interactions of primary cosmic rays with atmospheric nuclei~\cite{1991crpp.bookG}. 
Fluctuations in the atmospheric temperature and density contribute to variations in the detected muon rate. High energy muons can penetrate to deep underground laboratories if their energy is above a threshold value, $E_{\text{thr}}$, that depends on the depth level; lower energy muons are absorbed in the rock overburden. Numerous underground detectors have observed an annual modulation of the rates for high energy muons~\cite{Gaisser:2011cc}, including MACRO~\cite{Ambrosio:1997tc}, LVD~\cite{Aglietta:1998nx,Vigorito:2017xbb}, BOREXINO~\cite{Bellini:2012te,Agostini:2018fnx}, GERDA~\cite{Agostini:2016gda}, OPERA~\cite{Agafonova:2018kce}, IceCube~\cite{Desiati:2011hea}, MINOS~\cite{Adamson:2009zf,Adamson:2014xga}, Double CHOOZ~\cite{Abrahao:2016xio}, and Daya Bay~\cite{An:2017wbm}. 
There was only one reported search for daily variations of muon rates in an underground laboratory by the MACRO experiment in Granssaso (LNGS) with a modulation amplitude limit of less than 0.1\%~\cite{PhysRevD.67.042002}. 

However, the Tibet air shower array experiment (an above-ground experiment) observed a significant diurnal modulation ($0.5\%$) using an order of 10~TeV cosmic-rays~\cite{Munakata:869497}; the Aragats Multichannel Muon Monitor at high altitude (3200m) observed a similar level of diurnal modulation of the muon rate with energies higher than 5~GeV/c$^2$~\cite{MAILYAN20101380}.
These variations are of particular interest for the study of candidate for dark matter particles that might have a diurnal modulation~\cite{Hasenbalg:1997hs,Kouvaris:2014lpa,Kavanagh:2016pyr}. It is also important in understanding possible systematic effects for directional searches of the dark matter~\cite{Grothaus:2014hja} that search for daily modulation of WIMP directions caused by the daily rotation of the Earth. 

The COSINE-100 experiment's aim is to confirm or refute the annual modulation signal observed by the DAMA/LIBRA experiment with an array of 106 kg of low-background NaI(Tl) crystals at the Yangyang underground laboratory~(Y2L) in South Korea~\cite{Adhikari:2017esn,Adhikari:2017gbj}. Physics data have been collected since September 2016, and used to produce a series of first results and bounds on various dark matter models~\cite{Adhikari:2018ljm,Adhikari:2019off,Ha:2018obm,Adhikari:2019tgv}.  The first measurement of an annual event rate modulation was reported in 2019~\cite{Adhikari:2019off} even though its sensitivity was not sufficient to probe the DAMA/LIBRA signal. In this model independent measurement, it is crucial to understand the annual modulation of every environmental parameter, especially the cosmic-ray muon rate, as precisely as possible. 

In this article, we present an analysis of the cosmic muon flux measured by the COSINE-100 detector based on 952 live days data. An annual modulation of the muon rate at Y2L is reported for the first time. A study of correlations between the muon rate and the atmospheric temperature variations are in good agreement with the standard meson production in the atmosphere. 

\section{Muon flux at Y2L}
\label{sec:cosine_100}
\begin{figure}[htbp]	
	\centering
	\includegraphics[width = 0.9 \columnwidth] {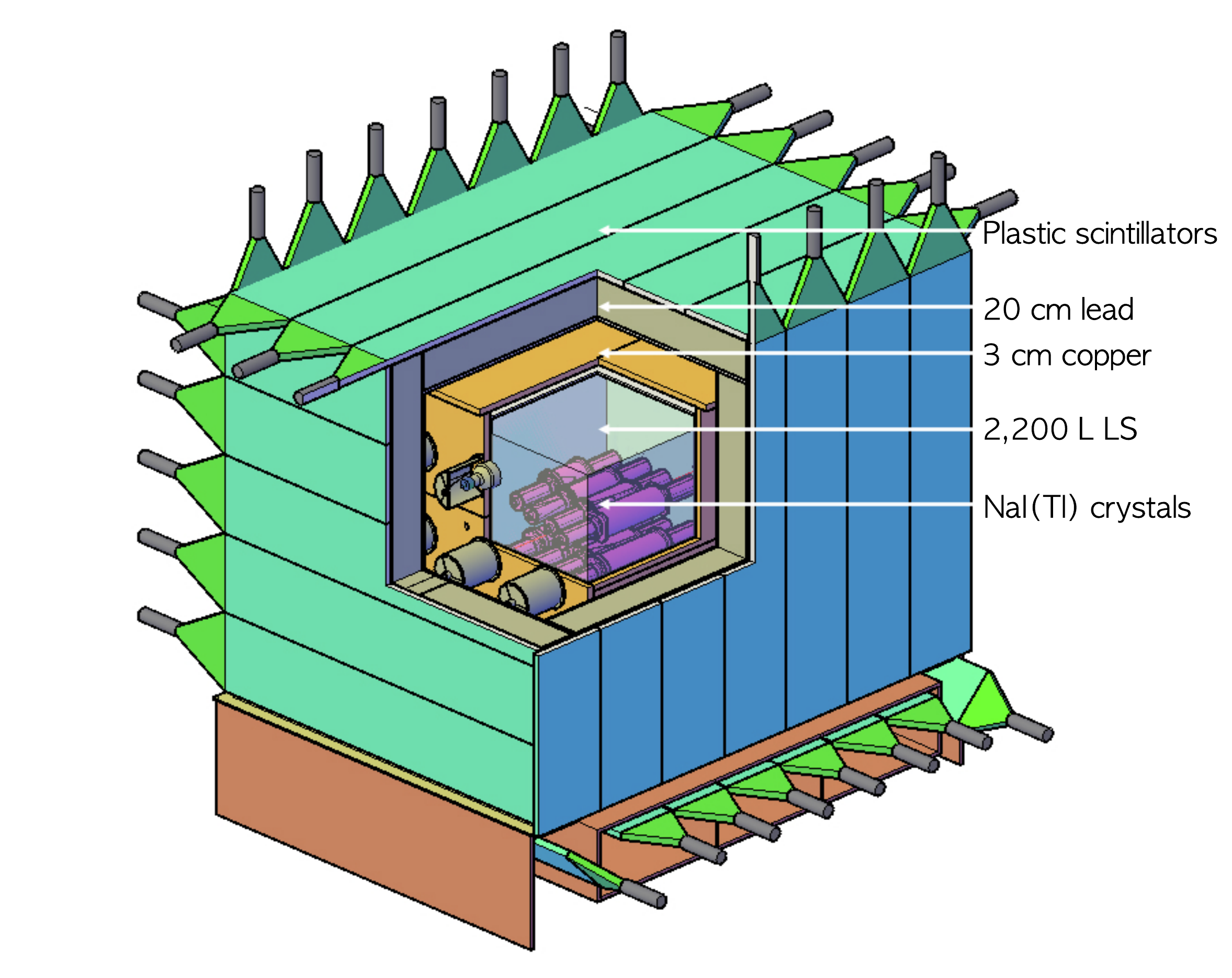}
	\caption{A schematic view of COSINE-100 detector.}
	\label{detector}
\end{figure}
The COSINE-100 detector~\cite{Adhikari:2017esn} consists of eight low-background NaI(Tl) crystals~\cite{Adhikari:2017gbj}, arranged in a 4~$\times$~2 array that are immersed in a liquid scintillator provides the identification and subsequent reduction of radioactive backgrounds in the crystals~\cite{Park:2017jvs}. 2,200~L of linear alkylbenzene (LAB) based liquid scintillator is contained in a 1~cm thick acrylic box that is supported with a 3~cm thick oxygen-free copper box. This is surrounded by a 20~cm thick lead bricks and 3-cm thick plastic scintillator panels as shown in Fig.~\ref{detector}. 

The outermost part of the COSINE-100 shield consists of 37 panels that are made of 3-cm-thick Eljen EJ-200 plastic scintillator\footnote{http://www.eljentechnology.com} for detection of muon-induced signals as described in Ref.~\cite{Prihtiadi:2017inr}. Muon fluxes in underground laboratories are significantly reduced by ranging out the muons in the rock overburden. However, the most energetic muons can penetrate to the underground laboratory and pass through the detector materials where they produce large energy depositions. Muon energy deposits are typically greater than the minimum ionization energy, which is approximately 4.5~MeV for a 3-cm-thick plastic scintillator~\cite{Tanabashi:2018oca}. This is a much higher energy than the energies of typical $\gamma$ or $\beta$ environmental background components. COSINE-100 has a 4$\pi$ muon detector that surrounds the entire shield so that the muon events can be identified by requiring hits on at least two sides of the detector array. Muon events can be selected by applying threshold requirements on deposited energy combined with limits on time correlations. A time difference requirement ($\Delta T$) is established that covers a 5$\sigma$ range of signal events, as described in Ref.~\cite{Prihtiadi:2017inr}. Most non-muon events with low energy deposits are rejected by the application of these requirements, with a loss in efficiency that is almost negligible.

Reflecting the cubic structure of the shield, we define the muon detector as having six sides~(top, bottom, front, back, left, and right). For the muon flux measurement, muon-candidate events passing through the top-side of the muon detector are used. A muon candidate event has a hit in the top-side plus one of the other five sides in coincidence. The effective area of the top-side, (5.48 $\pm$ 0.16) m$^{2}$, is used to determine the normalized muon rate. With data obtained between September 2016 and July 2019, corresponding to 952~days, we measured the average muon flux ($I_{\mu}^{0}$) at Y2L as 37.95~$\pm$~0.03$_\text{stat.}$~$\pm$~1.10$_\text{syst.}\times10^{-4}$muons/(s$\cdot$ m$^{2}$) where the systematic error is due to uncertainties in the effective areas of the muon counters that are not fully active. This number is consistent with a previous measurement based on the initial three month data-taking. Details of the muon flux measurement are described elsewhere~\cite{Prihtiadi:2017inr}.

\section {Annual modulation of the muon flux at Y2L}
\subsection{The muon flux variation}
\label{sec:modulation}
The meson production rate is affected by seasonal temperature variations in the upper atmosphere. Year-scale temperature modulations mainly occur due to the varying solar exposure through the year, such that the higher temperature in the summer lowers the average density which increases the density in winter, thereby altering the mean free path of the produced mesons. Only muons with high energy (greater than energy threshold $E_\text{thr}$) can penetrate to Y2L and these are consisting most likely the decay products of charged pions ($\pi^{\pm}$); only a small fraction from kaons ($K^{\pm}$). 
As a consequence, the cosmic muon flux as measured by COSINE-100 is expected to follow the modulation of the atmospheric temperature. At first order, the muon flux $I_{\mu}$(t) can be described by the simple sinusoidal form
\begin{equation}
I_{\mu}(t)=I_{\mu}^{0}+\Delta I_{\mu}~=~I_{\mu}^{0}+{\delta}I_{\mu}~\text{cos} \Bigl( \frac{2\pi}{T}(t-t_{0})\Bigr),
\label{eq:muonfit}
\end{equation}
where $I_{\mu}^{0}$ is the mean muon flux, $\delta I_{\mu}$ is the modulation amplitude, $T$ is the period, and $t_{0}$ is the phase of the modulation. 

\begin{figure*}[htbp]	
	\centering
	\includegraphics[width = 1.1\textwidth] {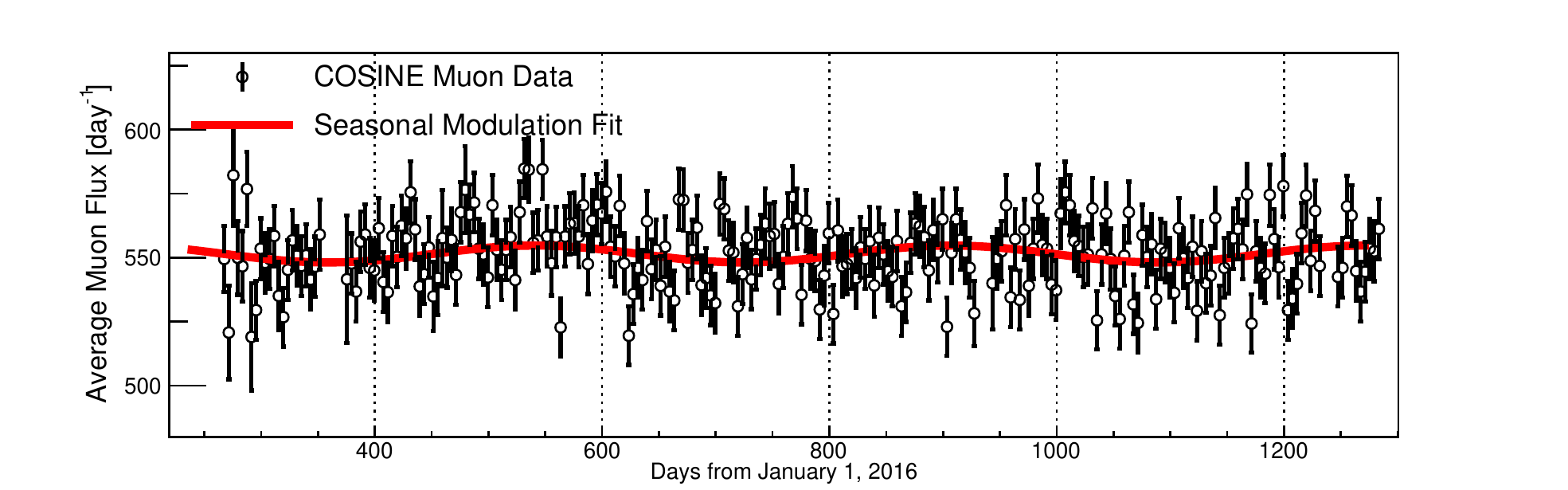}
	\caption{Cosmic muon rates at Y2L measured by the COSINE-100 detector as a function of time in 4-days bin is shown in points. The red line is the sinusoidal fit to the data assuming a seasonal modulation. 
	}
	\label{muonmodulation}
\end{figure*}

The muon rate ($I_{\mu}^{\text{0}}$) at Y2L has been measured over a three years period as described in Sec.~\ref{sec:cosine_100}. To determine its the correlation with the atmospheric temperature behavior correctly,  we only use muon events passing through the top- to the bottom-side panels that most likely have near-vertical trajectories at Y2L.
The measured  muon rates using the top-bottom coincidence condition as a function of time is shown in Fig.~\ref{muonmodulation}. The data are fitted with the sinusoidal function of Eq.~\ref{eq:muonfit} that is overlaid as a red solid line. In this fit, we assume a 1 year period~($T=1$~year). An average muon intensity $I_{\mu}^{0}=$~551.43 $\pm$ 0.78~muons$\cdot\text{day}^{-1}$ and a phase $t_{0}$ = (179 $\pm$ 19) days, corresponding to a maximum on the $27^{th}$ of June, are obtained with a goodness of fit $\chi^{2}/\text{NDF}=287.4/240$. A modulation amplitude of $\delta I_{\mu}$~=~3.31 $\pm$ 1.09~muons$\cdot\text{day}^{-1}$, corresponding to $\Delta I_{\mu}/I_{\mu}^0 =$~0.60 $\pm$ 0.20~\%, is obtained. This is the first measurement of seasonal modulation of the muon rate at Y2L. 

\subsection{The atmospheric model}
\label{sec:atmospheric}
The amplitude of the muon modulation (${\Delta I_{\mu}}/{I^{0}_{\mu}}$) is related to variations of the atmospheric temperature and pressure at various altitudes. The atmosphere is modeled as an isothermal gas for $\pi^{\pm}~\text{and}~K^{\pm}$ meson productions  with nuclei. The dependence of muon flux variations on the atmospheric temperature can be phenomenologically expressed as~\cite{RevModPhys.24.133}:
\begin{equation}
\label{eq:intensity1}
\frac{\Delta I_{\mu}}{I^{0}_{\mu}} = \int_{0}^{\infty}dX\alpha(X)\frac{\Delta T(X)}{T(X)},
\end{equation}
where $I^{0}_{\mu}$ is the average muon intensity measured at Y2L and $\Delta I_{\mu}$ is fluctuation of muon flux; $\alpha(X)$ is the temperature coefficient that relates fluctuations in the atmospheric temperature at depth $X$, $\frac{\Delta T(X)}{T(X)}$, to the fluctuations in the muon intensity. The integral extends over atmospheric depth from the highest altitude of pion production to the ground. The change in the Y2L muon rate can be rewritten as~\cite{RevModPhys.24.133,Ambrosio:1997tc,Adamson:2009zf}:
\begin{equation}
\label{eq:intensity2}
\Delta I_{\mu} = \int_{0}^{\infty}dXW(X)\Delta T(X),
\end{equation}
where the weight $W(X)$ reflects the temperature dependence of the production of mesons in the atmosphere and their decay into muons that are observed in the Y2L muon detector. From Eq.~\ref{eq:intensity1} and \ref{eq:intensity2}, a relation between the temperature coefficient $\alpha(X)$ and the weight $W(X)$ is:
\begin{equation}
\label{eq:coeffi}
\alpha(X)=\frac{T(X)}{I_{\mu}^{0}} W(X). 
\end{equation}

The atmosphere can be described by many layers with a continuous variation of temperature and pressure. A possible parametrization considers the atmosphere as a body of isothermal layers with pressure $X_{n}$ and temperature $T(X_{n})$ and defines an effective temperature, $T_{\text{eff}}$, as the weighted average over the atmospheric depth~\cite{Adamson:2009zf}:
\begin{equation}
\label{eq:teff}
T_{\text{eff}}\simeq \frac{\sum_{n=0}^{N}\Delta X_{n}T(X_{n})(W_{\pi}(X_{n}) + W_{K}(X_{n}))}{\sum_{n=0}^{N} \Delta X_{n} (W_{\pi} (X_{n}) + W_{K} (X_{n}))}.
\end{equation}
Here, $\Delta X_{n}$ is the difference between two adjunct pressure levels, and $W_{\pi,K}$ the weighting functions of the contributions of pions and kaons to the altitude dependence of the muon production, given by the following expression~\cite{Adamson:2009zf,Grashorn:2009ey}:
\begin{equation}
\label{eq:teff6}
W_{\pi,K}(X) \simeq \frac{(1-X/\Lambda_{\pi,K}^{'})^2 e^{-X/\Lambda_{\pi,K}} A_{\pi,K}^1}{\gamma+(\gamma+1)B_{\pi,K}^1 K_{\pi,K}(X) (\langle E_{\text{thr}}\text{cos}\theta\rangle/\epsilon_{\pi,K})^2},
\end{equation}
where,
\begin{equation} 
\label{eq:Kfactor}
K_{\pi.K}(X)~=~\frac{(1-X/\Lambda_{\pi,K}^{'})^2}{(1-e^{-X/\Lambda_{\pi,K}^{'}})\Lambda_{\pi,K}^{'}/X}.
\end{equation}
The parameters $A_{\pi,K}^1$ describe the relative contribution of kaons/pions and include the flux of inclusive mesons, the masses of mesons and muons, and the muon spectral index $\gamma$. The parameters $B_{\pi,K}^1$ reflect the relative atmospheric attenuation of mesons. The threshold energy, $E_\text{thr}$, is the minimum energy required for a muon to penetrate to Y2L depth. The parameters $\Lambda_{\pi,K}$ are pion and kaon attenuation lengths: $1/\Lambda_{\pi,K}^{'}=1/\Lambda_{N}-1/\Lambda_{\pi,K}$, where  $\Lambda_{N}$ is attenuation length of the primary cosmic ray.  The input parameters are taken from Refs.~\cite{Adamson:2009zf,Agostini:2018fnx,Grashorn:2009ey}. Since $E_{\text{thr}}$ depends on the rock overburden that a muon must penetrate to arrive at Y2L, the mean of the product of the threshold energy and the cosine of the zenith angle, $\langle E_{\text{thr}}\text{cos}\theta\rangle = 795 \pm 140~\text{GeV}$, which is corresponds to Y2L, is used for $T_{\text{eff}}$ determination. 

\begin{figure*}[htbp]	
	\centering
	\includegraphics[width = 0.9\textwidth] {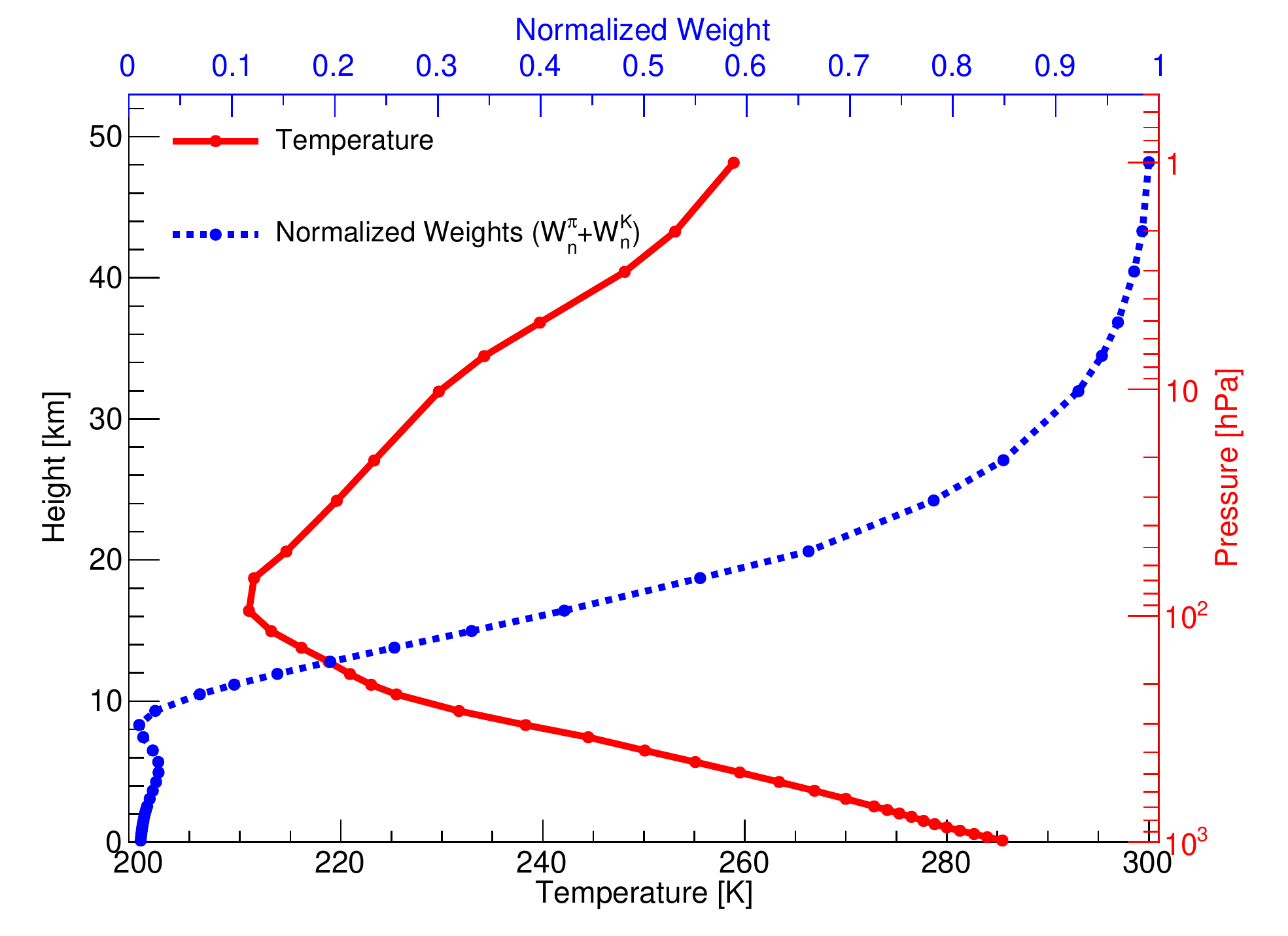}
	\caption{ The 952 days average temperatures at the location of Y2L at different heights are shown as the red solid-line and the normalized weighting factor $W_{n}^{\pi}+W_{n}^{K}$ as the blue dashed-line, as functions of the pressure levels. The left vertical axis shows the altitude corresponding to the pressure level on the right vertical axis.
	}
	\label{teff}
\end{figure*}

Figure~\ref{teff} shows the average temperature variations at different height levels used in for our three year running period data for the closest vertical point to Y2L, provided by the European Center for Medium-range Weather Forecasts~(ECMWF)~\cite{ecmwf} and normalized weight factors to the respective altitude levels. Because muons are mainly produced at higher altitude, the higher layers of the atmosphere are assigned higher weights. 

\begin{figure*}[htbp]	
	\centering
	\includegraphics[width = 1.1\textwidth] {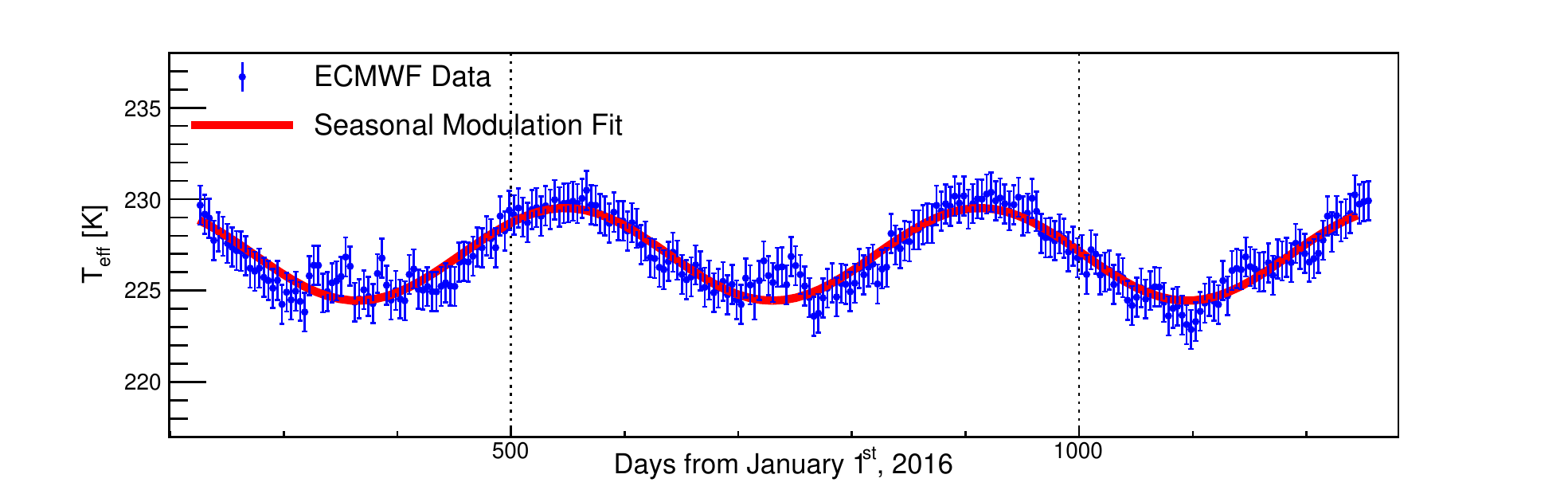}
	\caption{Effective atmospheric temperature at Y2L location calculated from Eq.~\ref{eq:sinusoidal_teff} as a function of time in a 4-days bin is shown as points. The red solid-line is the sinusoidal fit to the data.
	}
	\label{teffy2l}
\end{figure*}

The temperature data from ECMWF exploits different types of observations (e.g. surface, satellite, and upper air sounding) at many locations around the world, and uses a global atmospheric model to interpolate to Y2L. We collected temperature data in the same location of Y2L for 37 discrete altitude levels in the [0-50]~km range (Fig.~\ref{teff} solid-line), four times a day at 00.00 h, 06.00 h, 12.00 h, and 18.00 h GMT time. Based on this data set as well as the calculated altitude-dependent weight factors~(Fig.~\ref{teff} dashed-line), $T_{\text{eff}}$ values were calculated four times per day and averaged for daily data.  As one can see in Fig.~\ref{teffy2l}, $T_{\text{eff}}$ as a function of time with 4-day bins over a three-year interval shows a clear seasonal variation. At first order, the effective temperature $T_{\text{eff}} $ can be described by a simple function,

\begin{equation}
\label{eq:sinusoidal_teff}
T_{\text{eff}}(t)~=~T_{\text{eff}}^{0}+\Delta T_{\text{eff}}~=~T_{\text{eff}}^{0}+\delta T_{\text{eff}}~\text{cos} \left (\frac{2\pi}{T}(t-t_{0})\right),
\end{equation}
where $T_{\text{eff}}^{0}$ is the average effective temperature, $\delta T_{\text{eff}}$ is the modulation amplitude, $T$ is the period, and $t_0$ is the phase.  

The fit to the  $T_{\text{eff}}$ data is overlaid in Fig.~\ref{teffy2l}. Similarly to the muon flux data, we assume a 1 year period. An average effective temperature $T_{\text{eff}}^{0}$~=~229.0 $\pm$ 0.1 K and a phase $t_{0}$ = (183 $\pm$ 2) days, which is corresponding to a maximum on the $1^{st}$ of July, with goodness of fit  $\chi^{2}/\text{NDF}$=124.9/253 are obtained.  The measured phase of  $T_{\text{eff}}^{0}$ is consistent with the phase of the muon rate at Y2L.  The fitted modulation amplitude is $\delta T_{\text{eff}}=2.43 \pm 0.13$ K corresponding to $\Delta T_{\text{eff}}/T_{\text{eff}}^{0} = 1.06 \pm 0.05$ \%. 

\subsection{Correlation coefficient}

\begin{figure*}[htbp]	
	\centering
	\includegraphics[width = 1.1\textwidth] {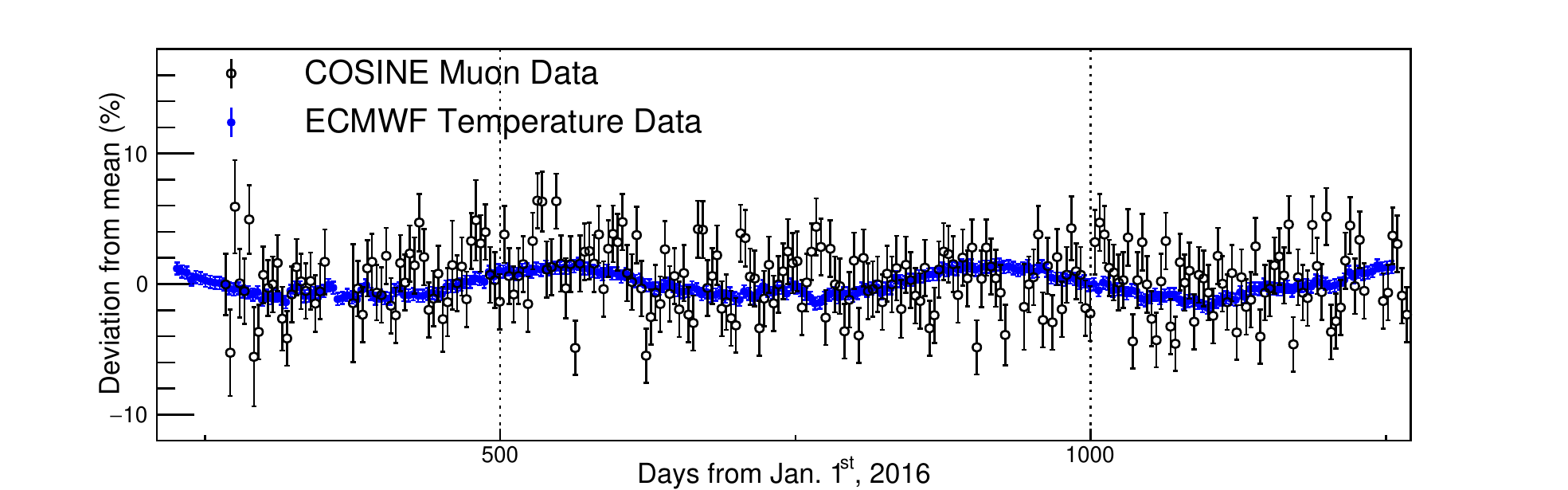}
	\caption{Variations of the  muon rate and the effective atmospheric temperature at Y2L. 
	}
	\label{devmodulation}
\end{figure*}

The muon rate and the effective temperature can be characterized by an effective temperature coefficient~($\alpha_T$) using Eq.~\ref{eq:intensity2}~and~\ref{eq:coeffi},
\begin{equation}
\alpha_{T}~=~\frac{T_{\text{teff}}^{0}}{I_{\mu}^{0}}~\int_{0}^{\infty} dXW(X),
\end{equation}
where $W(X)=W_{\pi}(X)+W_{K}(X)$. We also simplify Eq.~\ref{eq:intensity1} as,
\begin{equation}
\frac{\Delta I_{\mu}}{I_{\mu}^{0}}~=~\alpha_{T}~\frac{\Delta T_{\text{eff}}}{T_{\text{eff}}^{0}}. 
\label{eq:effective_temperature}
\end{equation}

\begin{figure}[htbp]	
	\centering
	\includegraphics[width = 0.9 \columnwidth] {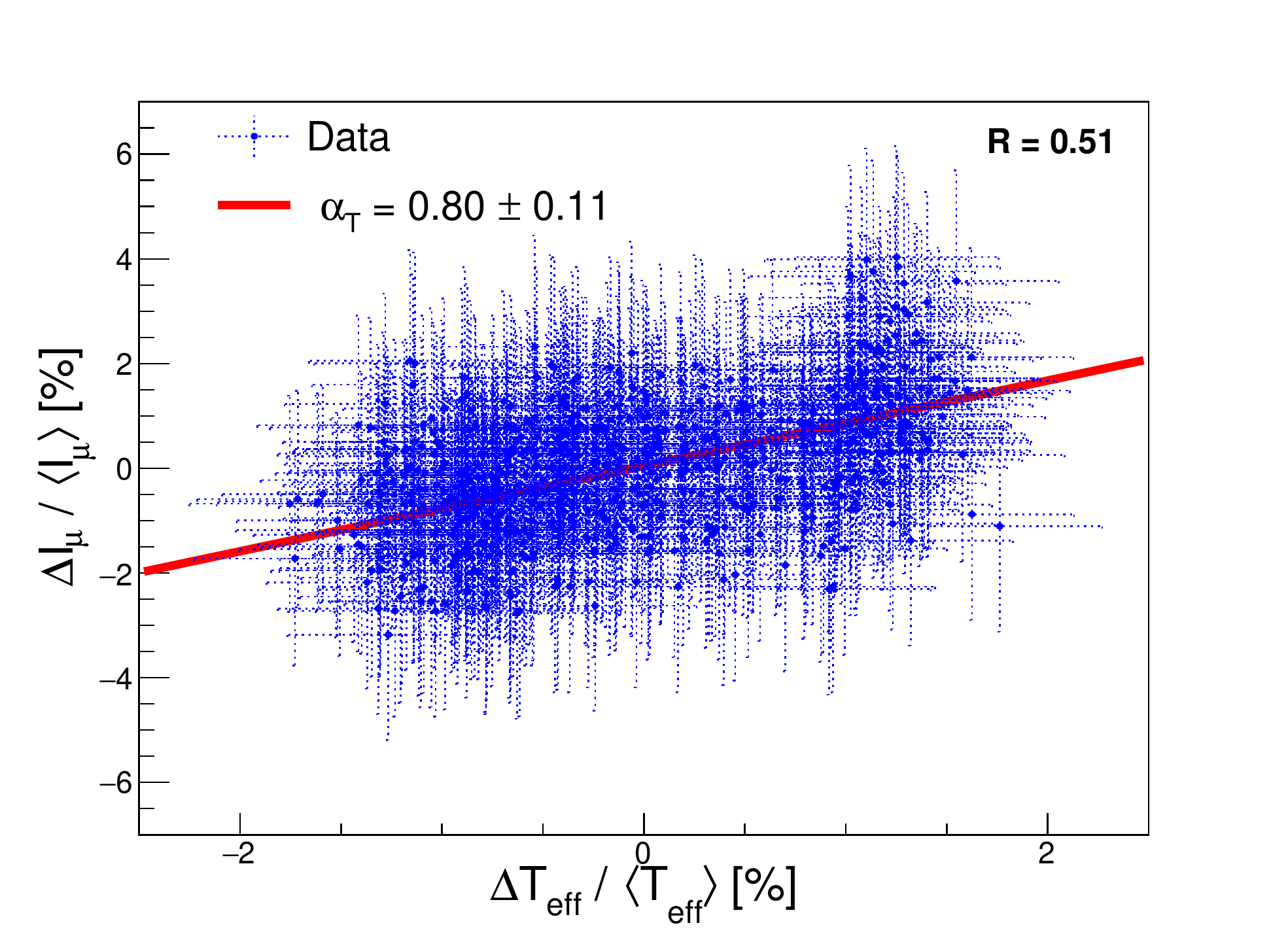}
	\caption{$\Delta I_{\mu}/I_{\mu}^{0}$ versus  $\Delta T_{\text{eff}}/T_{\text{eff}}^{0}$ where each point corresponds to a single day.}
	\label{alphaT_modulation}
\end{figure}

Figure~\ref{devmodulation} shows a comparision between the muon flux and the effective temperature that are scaled to percent deviations from their means $I_{\mu}^{0}$ and $T_{\text{eff}}^{0}$ and combined to 4-day bins. As expected from the previous discussion, the modulation parameters inferred for the cosmic muon rate and the effective temperature have a clear correlation.  To quantify this correlation, we plot $\Delta I_{\mu}/I_{\mu}^{0}$~vs~$\Delta T_{\text{eff}}/T_{\text{eff}}^{0}$ values for each day in Fig.~\ref{alphaT_modulation}. A fit with a linear function provides a correlation coefficient $\alpha_T=0.80 \pm 0.11_{\text{(stat.)}}$ between two parameters. The correlation coefficient (R-value) between these two distributions is 0.51 indicating a positive correlation. This is the first direct measurement of $\alpha_T$ at Y2L. 

\begin{figure}[htbp]	
	\centering
	\includegraphics[width = 0.9 \columnwidth] {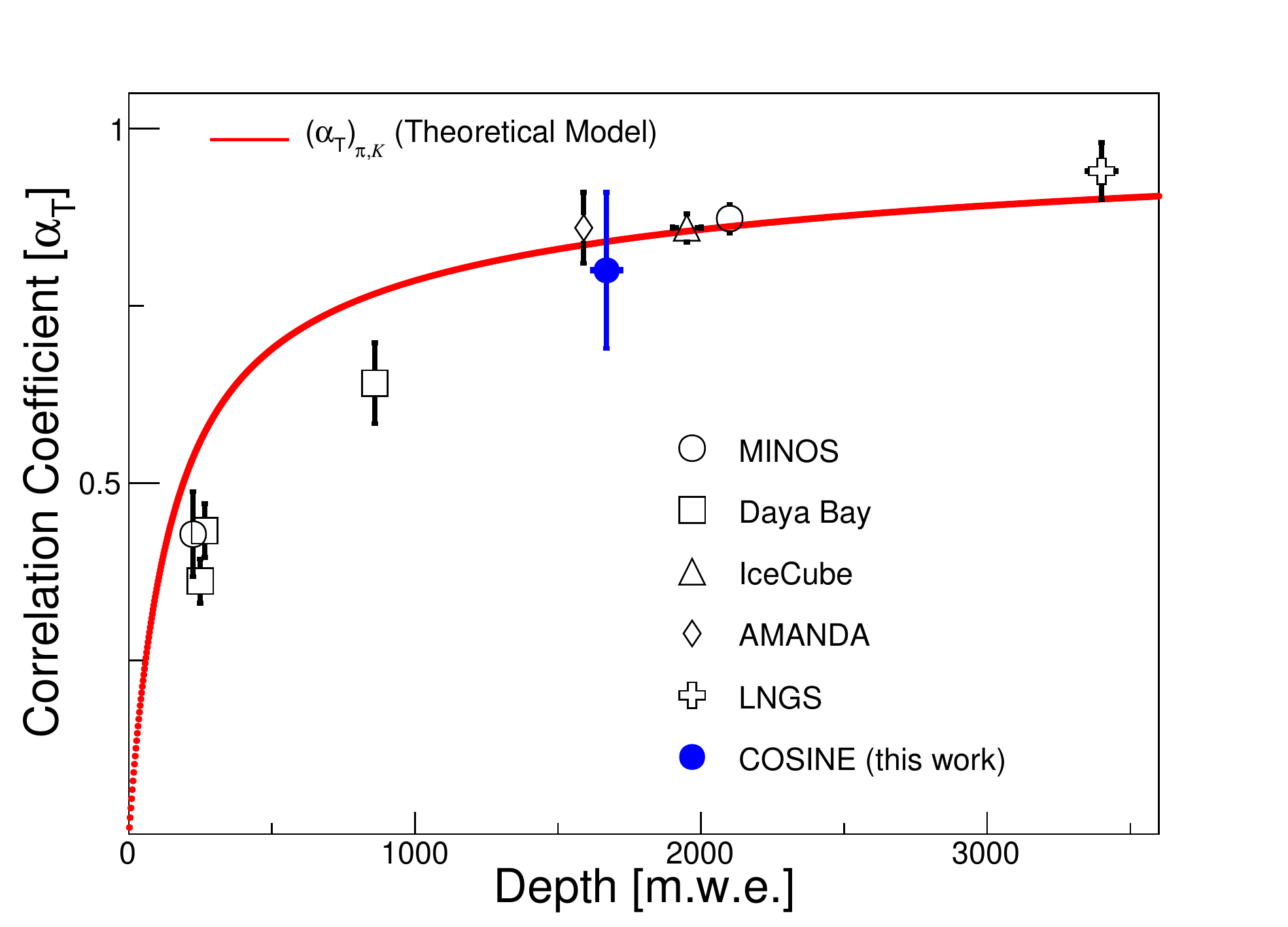}
	\caption{Measured $\alpha_{T}$ values as a function of the underground laboratories depth is shown.  Various measurements from different experiments are listed: MINOS~\cite{Adamson:2014xga}, Daya Bay~\cite{An:2017wbm}, IceCube~\cite{Desiati:2011hea}, AMANDA~\cite{Bouchta:1999kg}, and LNGS together with our Y2L measurement~(blue filled circle). Here, the LNGS measurement is a weighted average of various Granssaso experiments taken from Ref.~\cite{Agostini:2018fnx}. The red solid line show predictions based on a meson production models for a reasonable atmospheric kaon/pion ratio~\cite{1991crpp.bookG,Gaisser:2011cc}.}
	\label{effectivetemp}
\end{figure}

The values of $\alpha_T$ as a function of observation depth can be used to test the meson production model as shown in Fig.~\ref{effectivetemp}. A model calculation with the kaon-pion production ratio $r_{K/\pi} = 0.149 \pm 0.06$~\cite{1991crpp.bookG,Gaisser:2011cc} is shown as a red solid line. Our measurement of $\alpha_T$ in Y2L is in good agreement with the predictions of the model. 

\section{Diurnal modulation}

\begin{figure}[htbp]	
	\centering
	\includegraphics[width = 1.0 \columnwidth] {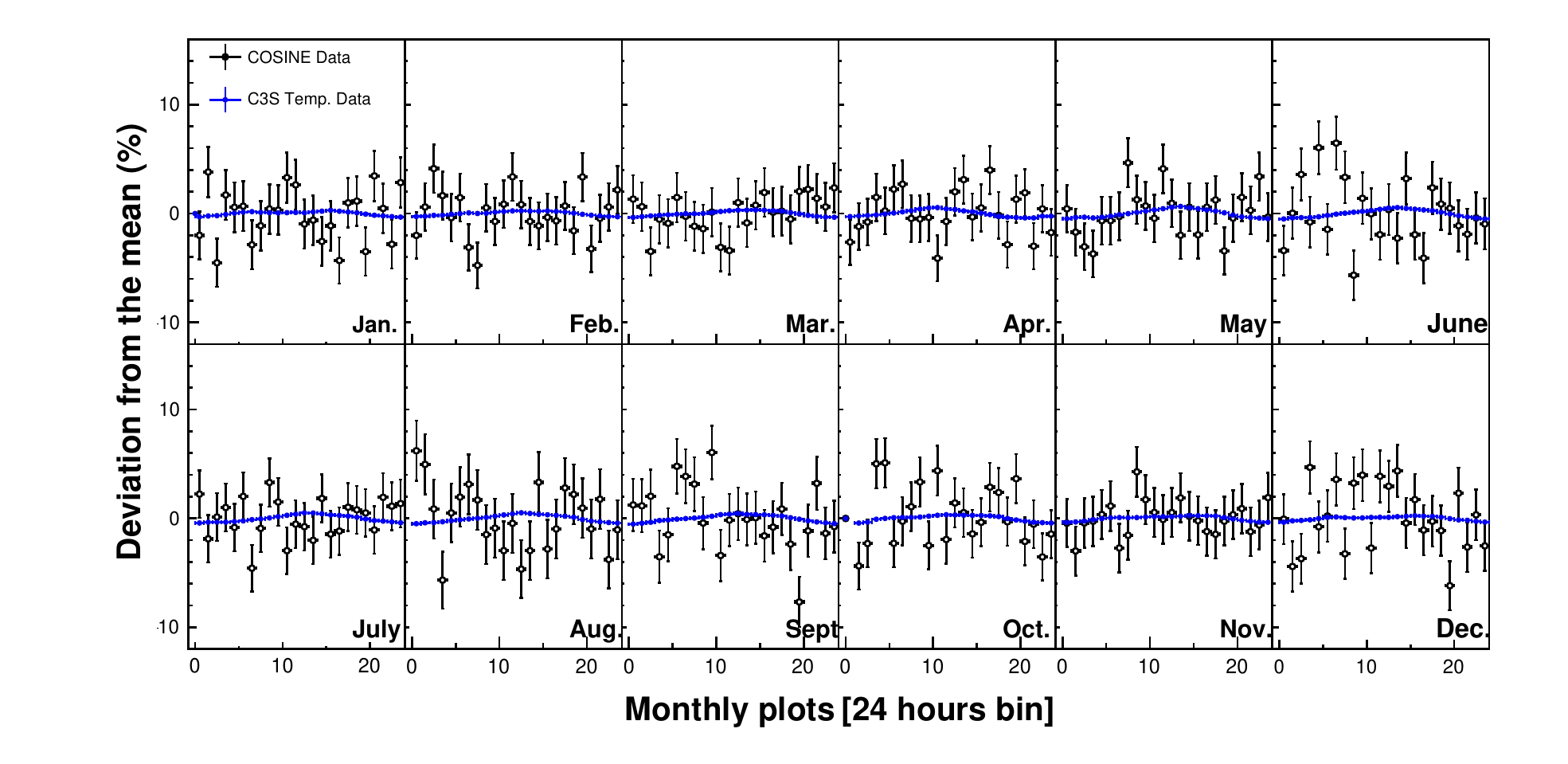}
	\caption{The variations of the muon rate and effective temperature in 1 hour bins according to the solar local diurnal time at Y2L are shown for 12 calendar months. The data are average values of a same month averaged over three years.}
	\label{devmonthly}
\end{figure}

We also conducted a study on daily muon flux variations. We averaged out all data in the same month of the year from Jan. to Dec. into 24 daily 1-hour bins for seasonal effects of diurnal variations. The 24 hour muon rate variations for the averaged data for each month are shown in Fig.~\ref{devmonthly} as percent deviations from the mean. The modulations of the effective temperature in the same bin size are overlaid, where the atmospheric temperature data were retrieved from the Copernicus Climate Change Service (C3S)~\cite{era5} that provides temperature for every hour near Y2L location. No statistically significant diurnal modulations are observed.

\begin{figure}[htbp]	
	\centering
	\includegraphics[width = 0.7 \columnwidth] {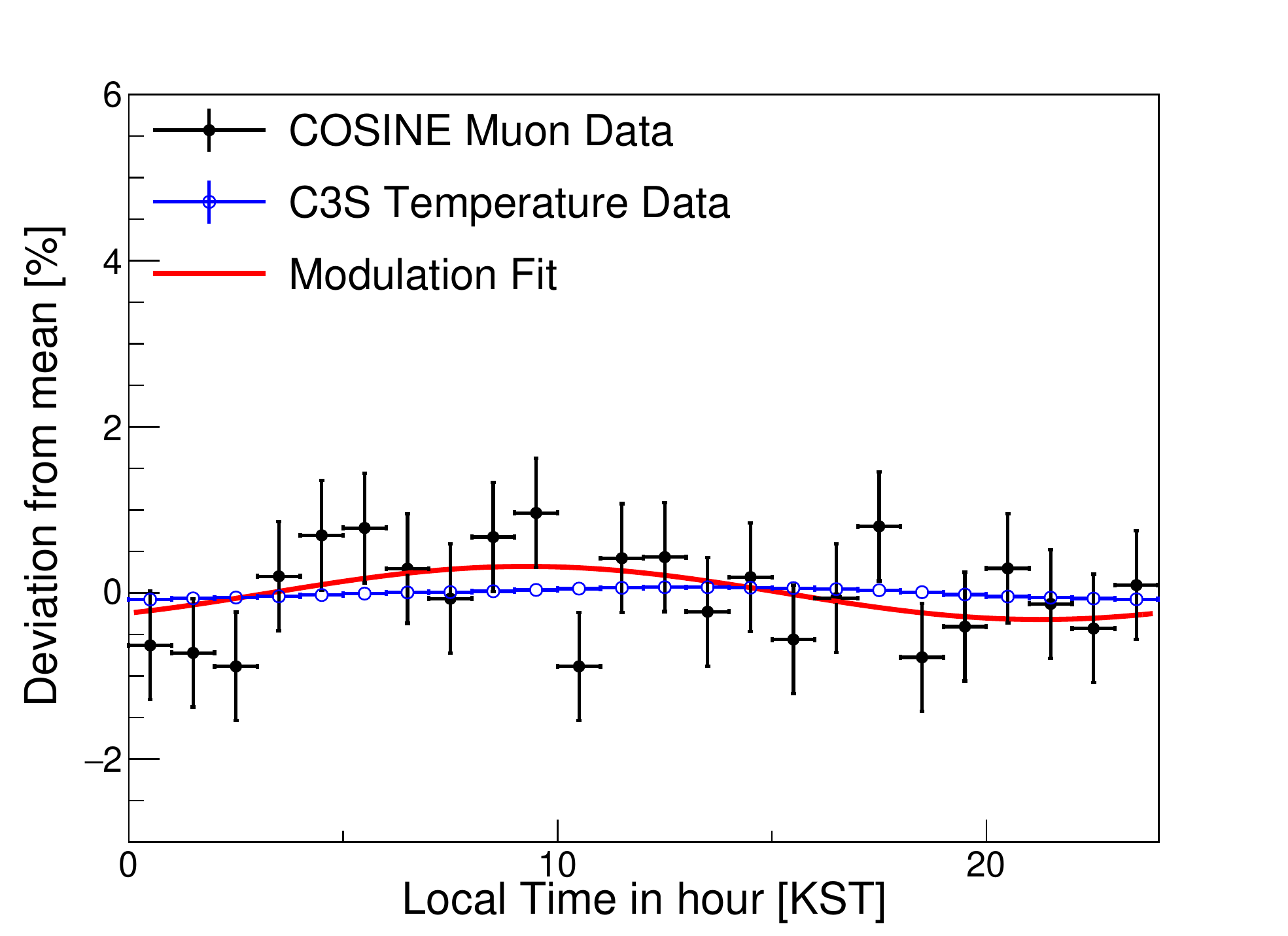}
	\caption{Daily variations of the muon and temperature-corrected hourly data fitted to a cosine function. The red-line is the best-fit curve according to the form Eq.~\ref{eq:muonfit} representing the muon rate modulation.}
	\label{devdaily}
\end{figure}

We averaged all 12-months data into one plot for the muon rate as well as the effective temperature in a 1 hour bins as shown in Fig.~\ref{devdaily}. We fit the muon data using a sinusoidal function and observed a 0.32 $\pm$ 0.19\% modulation amplitude. 
This result is consistent with the 0.3$~-~$0.5\% modulation observed by above ground measurements of the muon and cosmic rays~\cite{MAILYAN20101380,Munakata:869497}, but also consistent with the MACRO experiment's of $<$0.1\% upper limit on the diurnal modulation amplitude~\cite{PhysRevD.67.042002}. 

\section{Conclusions}
\label{sec:conclusions}
We report first measurements of an annual modulation  as well as a limit on the diurnal modulation of the muon rate at Y2L. The fractional annual modulation amplitude is measured to be $\Delta I_{\mu}/I^{0}_{\mu}$=0.60 $\pm$ 0.20\% (a maximum phase at 179 $\pm$ 19 day starting from Jan 1$^{st}$, corresponding to June 27$^{th}$). The effective temperature and muon production model describe the observed annual modulation data very well. We found no significant modulation of the diurnal muon rate. 

\acknowledgments
We thank the Korea Hydro and Nuclear Power (KHNP) Company for providing underground laboratory space at Yangyang. This work is supported by: the Institute for Basic Science (IBS) under project code IBS-R016-A1 and NRF-2016R1A2B3008343, Republic of Korea; NSF Grants No. PHY-1913742, DGE-1122492, WIPAC, the Wisconsin Alumni Research Foundation, United States; STFC Grant ST/N000277/1 and ST/K001337/1, United Kingdom; and Grant No. 2017/02952-0 FAPESP, CAPES Finance Code 001, CNPq 131152/2020-3, Brazil.

\bibliographystyle{JHEP}
\bibliography{dm}

\end{document}